# Cross-National Measurement of Polarization in Political Discourse

Analyzing floor debate in the U.S. and the Japanese legislatures

Takuto Sakamoto
Graduate School of Arts and Sciences
University of Tokyo
Tokyo, Japan
sakamoto@hsp.c.u-tokyo.ac.jp

Hiroki Takikawa
Frontier Research Institute for Interdisciplinary Sciences
Tohoku University
Sendai, Japan
takikawa@m.tohoku.ac.jp

*Abstract*—Political polarization in public space can seriously hamper the function and the integrity of contemporary democratic societies. In this paper, we propose a novel measure of such polarization, which, by way of simple topic modelling, quantifies differences in collective articulation of public agendas among relevant political actors. Unlike most other polarization measures, our measure allows cross-national comparison. Analyzing a large amount of speech records of legislative debate in the United States Congress and the Japanese Diet over a long period of time, we have reached two intriguing findings. First, on average, Japanese political actors are far more polarized in their issue articulation than their counterparts in the U.S., which is somewhat surprising given the recent notion of U.S. politics as highly polarized. Second, the polarization in each country shows its own temporal dynamics in response to a different set of factors. In Japan, structural factors such as the roles of the ruling party and the opposition often dominate such dynamics, whereas the U.S. legislature suffers from persistent ideological differences over particular issues between major political parties. The analysis confirms a strong influence of institutional differences on legislative debate in parliamentary democracies.

I. INTRODUCTION

Collective decision-making through public deliberations constitutes an essential component of democracy. Heated arguments naturally take place in these deliberations. Yet, the underlying division in the public space sometimes rapidly and wildly widens, exceeding the level at which competing arguments can be reconciled with each other. This process of political polarization, which has been observed in numerous examples of parliamentary gridlock, can pose a serious threat to the proper working of democratic governance itself. For this reason, political polarization has attracted attention from a number of scholars, especially political scientists [1-6]. Some of these scholars have devised sophisticated quantitative measures of polarization and employed these measures to analyze a wide array of large empirical data: from detailed records of voting behavior of lawmakers inside the United States Congress to over 2 million books published over a century [4,5]. Indeed, the inquiry into polarization is a natural field in which computational social science and big data can make a huge contribution to existing social sciences [7].

In this paper, we propose a novel measure of political polarization. This measure is derived from topic modelling of text data. Topic models, which are widely-used tools in computational social science, have also been increasingly employed in the political science literature [8-12]. A topic model, if properly specified, can automatically extract latent semantic structure of the corpus concerned and reveal distribution of the attention of each author (speaker, writer…) of a text to the semantic groupings ('topics') thus extracted [13]. This fits the traditional interests of political scientists in subjects such as agenda setting and issue articulation [14-16], which often critically determine the direction of political discourse. By relying on topic modelling, our measure can capture polarization occurring at a deeper level than those revealed by, for instance, counting frequencies of specific 'partisan' words [4,6] or comparing roll-call voting patterns of lawmakers [5].

Moreover, our measure of political polarization has another advantage over other preceding measures. The latter typically assume the existence of a specific institutional arrangement (e.g., roll-call voting) or political configuration (e.g., ideological division between two parties such as the Democratic and Republican parties). These assumptions make it difficult to carry the derived quantities beyond a single-county context. In contrast, our measure, being constructed from the simplest topic model, unsupervised latent Dirichlet allocation (LDA) [13], requires no such assumptions (except for the availability of a sizable volume of political documents). The measure is easy to implement and amenable to cross-national comparison of political polarization.

In the study reported below, we actually apply this measure to political texts from two countries, the United States and Japan. Specifically, we analyze speech documents that record legislative debate in the United States Congress and the Japanese Diet. The data, which contains more than 180 million words in total and spans a period of more than two decades, far surpasses in its coverage the legislative speech records that have so far been analyzed by other researchers employing topic models [6, 12]. Transcending a single-country/cultural/societal context, which is still a rarity in the relevant literature [11], this study will help further broaden the scope of the growing field of computational social science at large.

## II. DATA AND METHODS

### A. Data

We obtained records of floor deliberations in the U.S. Congress from the much-used *Congressional Record*, whose online version is available from the website of the U.S. Government Publishing Office (GPO)[1]. The obtained records cover both the House of Representatives and the Senate during a period from January 1994 to December 2016. Since the original dataset consists of a series of unstructured plain-text files, we used a third-party parser[2] to convert these files into an XML format where subsidiary information such as the name of a speaker was separated and tagged. We also accessed *ProPublica Congress API*[3] in order to obtain other relevant information that cannot be fully known from *Congressional Record*, including the party affiliation of each congressperson.

We then applied a standard set of preprocessing procedures against the obtained speech records. Using the Natural Language Toolkit (NLTK) with Python (version 3.5.2), we split up ('tokenized') the entire data on a single-word basis ('uni-gram'), removed commonly-used 'stop words' (e.g., 'is', 's', 'am', 'or', 'who'), and stemmed each of the tokens with the widely-used Porter stemmer. The resulting 'bag of words' consists of 1,074,906 units of speech (mostly separated by the change of speakers) with a total of 182,796,018 tokens (177,330 unique tokens). When applying a topic model to this corpus, we further removed extremely frequent (appearing in more than 50% of speeches) and extremely infrequent (appearing less than 50 times in the entire corpus) tokens for the purpose of computational efficiency. As a result, the number of tokens was reduced to 100,119,630 (22,254 unique ones).

As to the Japanese Diet, we accessed an online version of *Diet Conference Proceedings*[4]. Employing Python's selenium package, we extracted the minutes of all the plenary sessions and all the major committees of both the House of Representatives and the House of Councilors that had been held between 1994 and 2016. We used a morphological analyzer (MeCab) to tokenize the speech data, and then removed a set of stop words. After further removing extreme-frequency words just as in the U.S. case, the final corpus was obtained. This bag-of-words consists of 2,230,363 units of speech, which contain a total of 83,040,813 tokens (38,302 unique ones).

### B. Topic Model

We suppose that the data generation process for these corpora is formally described by a simple topic model: latent Dirichlet allocation (LDA) [13]. LDA assumes the existence of a particular number ($K$) of latent semantic groupings (topics) behind a given corpus. Each document (i.e., a unit of speech in our case) $d$ (=1,…,$M$) that composes the corpus is then characterized by a particular probability mixture of these topics $\theta_d=(\theta_{d,1},…,\theta_{d,K})$, while each topic $k$ (=1,…,$K$) is substantiated by a probability distribution defined over $V$ words, $\phi_k=(\phi_{k,1},…,\phi_{k,V})$, that governs stochastic generation of each word $w$ ($\in \{1,…,V\}$). In LDA, these probability vectors are generated from the following two Dirichlet distributions.

$$\left. \begin{array}{l} \theta_d \sim Dirichlet(\alpha) \quad (d=1,\cdots,M), \\ \phi_k \sim Dirichlet(\beta) \quad (k=1,…,K). \end{array} \right\} \quad (1)$$

where $\alpha$ and $\beta$ are $K$-dimensional and $V$-dimensional vectors, respectively.

In this setting, a word $w_{d,i}$, which appears in the $i$th location of a document $d$, is stochastically generated through the following procedure. First, a topic is sampled from the multinomial distribution controlled by $\theta_d$. Denoting this topic with a latent variable $z_{d,i}$, the word in question, $w_{d,i}$, is chosen from $V$ words according to another multinomial distribution controlled by $\phi_{z_{d,i}}$. Thus,

$$\left. \begin{array}{l} z_{d,i} \sim Multinomial(\theta_d), \\ w_{d,i} \sim Multinomial(\phi_{z_{d,i}}). \end{array} \right\} \quad (2)$$

We trained this hierarchical model with the U.S. and the Japan speech corpora separately. In estimating the model's parameters such as $\theta_d$ and $\phi_k$, we used a library in Python's gensim package (gensim.models.ldamulticore.LdaMulticore), which employs computationally-efficient online variational Bayesian optimization as a learning algorithm [17]. For each country, we repeated the estimation over a range (30 to 80) of $K$ (the number of topics). Although we focus here on the case of $K$=70, the reported results, among them, the overall difference in polarization between the two countries, largely apply to the other cases at least qualitatively.

### C. Polarization Measure

The gensim library we used returns a stochastic estimate of a topic $z_{d,i}$ for each location $i$ of each document $d$ in the form of a probability distribution over $K$. By aggregating these estimates at a given level (individual, gender, party, chamber, country, etc.) over a given period of time, one can obtain a quantitative description of collective attention to the estimated topics on the specified scale. Let $\Theta_{G,\Delta T}$ denote this quantity for a unit of aggregation $G$ over a period $\Delta T$. $\Theta_{G,\Delta T}$ is essentially a composite probability distribution over $K$ topics where each of the aggregated documents is weighted with its word count.

Our measure of political polarization quantifies a difference between two such distributions. The difference is represented here as the Jensen-Shannon (JS) divergence between $\Theta_{G1,\Delta T1}$ and $\Theta_{G2,\Delta T2}$ (hereafter, shortened as $\Theta_1$ and $\Theta_2$). That is,

$$JS(\Theta_1,\Theta_2) = \frac{1}{2}\left(\sum_{k=1}^{K}\Theta_1(k)\log\frac{\Theta_1(k)}{\overline{\Theta}(k)}\right) + \frac{1}{2}\left(\sum_{k=1}^{K}\Theta_2(k)\log\frac{\Theta_2(k)}{\overline{\Theta}(k)}\right),$$
$$\overline{\Theta}(k) = \frac{1}{2}\Theta_1(k) + \frac{1}{2}\Theta_2(k) \qquad (3)$$

---

[1] https://www.gpo.gov/
[2] https://github.com/unitedstates/congressional-record
[3] https://propublica.github.io/congress-api-docs/
[4] http://kokkai.ndl.go.jp/

Note that the above polarization measure can be computed at any level of aggregation as far as the dimensions (i.e., the numbers of topics, $K_1$ and $K_2$) of $\Theta_1$ and $\Theta_2$ are the same. For example, the measure can capture diverging interests between two individual lawmakers, or it can track changing collective attention of an entire legislature between two time periods. In the following, we focus on the political party as a unit of aggregation. We first compare, between the U.S. and Japan, the degrees of inter-party polarization that are averaged over the entire period of investigation (1994-2016). This is followed by brief examination of their yearly changes.

### III. RESULTS

#### A. Cross-National Comparison

For both the U.S. and Japan, the estimated topic models generated mostly plausible semantic groupings. Tables I and II list 10 major topics (out of 70) in terms of their aggregated frequencies during 1994-2016 for the U.S. Congress and the Japanese Diet, respectively. Each topic is annotated with 10 most relevant tokens derived from estimated $\phi_k$. In the U.S. case, divisive policy issues such as public spending (topic 15) and health care (topic 8) are clearly identifiable among the listed topics. Similarly, the topics frequently debated in the Japanese Diet reflect persistent policy concerns of the country such as regional security situation (topic 31) and economic policies (topics 29 and 9). On the other hand, one should also notice that in either country these 'policy topics' are not the most dominant ones. In each legislature, the most frequent topic consists of general verbs and nouns that are typically uttered in the context of legislative debate (e.g., topic 29 in Table I and topic 24 in Table II).

The notable difference between the two countries emerges when one looks into political polarization among major parties. Fig. 1 and Fig. 2 illustrate this. Both compare topic distributions $\Theta_{G,\Delta T}$ that are aggregated at the party level for the entire period. In Fig.1, two dominant parties in the U.S. Congress, the Republican Party (conservative; red bars) and the Democratic Party (liberal; blue bars), are compared in their respective attention to 70 topics. Qualitatively, the two topic distributions show similar patterns. JS divergence, our measure of polarization, between these distributions is 0.008.

TABLE I. MAJOR TOPICS IN THE U.S. CONGRESS (1994-2016)

| Major Topics | Freq | Most Relevant Tokens |
|---|---|---|
| topic 29 | 0.089 | go, peopl, say, get, think, want, know, one, thing, talk |
| topic 33 | 0.033 | work, thank, bill, issu, import, senat, want, time, legisl, colleagu |
| topic 40 | 0.033 | year, --, one, famili, life, day, peopl, great, mani, man |
| topic 54 | 0.03 | vote, bill, republican, senat, democrat, --, pass, american, major, hous |
| topic 15 | 0.028 | budget, spend, debt, cut, year, deficit, --, tax, balanc, govern |
| topic 8 | 0.026 | health, care, insur, medicar, cost, plan, coverag, bill, senior, -- |
| topic 69 | 0.024 | year, percent, $, go, pay, 1, rate, billion, increas, cut |
| topic 65 | 0.024 | agenc, requir, secur, depart, act, report, commiss, administr, inform, govern |
| topic 23 | 0.021 | amend, would, offer, bill, time, chairman, languag, think, make, point |
| topic 12 | 0.021 | state, unit, nuclear, iran, weapon, china, nation, u, --, world |

TABLE II. MAJOR TOPICS IN THE JAPANESE DIET (1994-2016)

| Major Topics | Freq | Most Relevant Tokens |
|---|---|---|
| topic 24 | 0.051 | 出る(attend), 行く(go), やっぱり(as expected), 聞く(listen), 見る(see), けど(but), わかる(understand), 入る(enter), 来る(come), 使う(use) |
| topic 63 | 0.034 | 推進(promotion), 支援(support), 重要(important), 取組(work), 取り組む(work on), 進める(promote), 図る(plan), 確保(keep), 行う(carry out), 実施(implementation) |
| topic 28 | 0.034 | 法(law), 規定(provision), 条(article), 法律(law), 改正(amend), 定める(establish), 規制(regulation), 必要(necessary), 要件(requirement), 行う(carry out) |
| topic 20 | 0.032 | 国民(nation), 総理(PM), 改革(reform), 政権(administration), 政府(government), 政治(politics), 民主党(Democratic Party), 国会(Diet), 法案(bill), 皆様(everyone) |
| topic 50 | 0.029 | 検討(examination), 議論(discussion), 踏まえる(based on), 必要(necessary), 行う(carry out), 進める(promote), 含める(include), 具体(concrete), 見直し(review), 問題(issue) |
| topic 25 | 0.027 | 見る(see), 率(rate), 上がる(rise), 出る(appear), ふえる(increase), 高い(high), 減る(decrease), 意味(meaning), 議論(discussion), 上げる(raise) |
| topic 8 | 0.022 | 日本(Japan), 世界(World), 国(country), 社会(society), 歴史(history), 意味(meaning), 文化(culture), 大きな(large), 時代(time), 価値(value) |
| topic 31 | 0.021 | 日本(Japan), アメリカ(U.S.), 中国(China), 問題(issue), 総理(PM), 韓国(Korea), 外交(diplomacy), 外務大臣(foreign minister), 関係(relations), 北朝鮮(North Korea) |
| topic 29 | 0.021 | 中小企業(small firms), 事業(business), 経済(economy), 経営(management), 状況(situation), 資金(fund), 行う(carry out), 企業(firm), 支援(support), 運用(use) |
| topic 9 | 0.021 | 予算(budget), 財政(finance), 経済(economy), 円(Yen), 政府(government), 景気(business), 総理(PM), 削減(cut), 対策(measure), 国民(nation) |

The topic numbers indicate no sematic relationships between the corresponding topics in the U.S. and Japan.

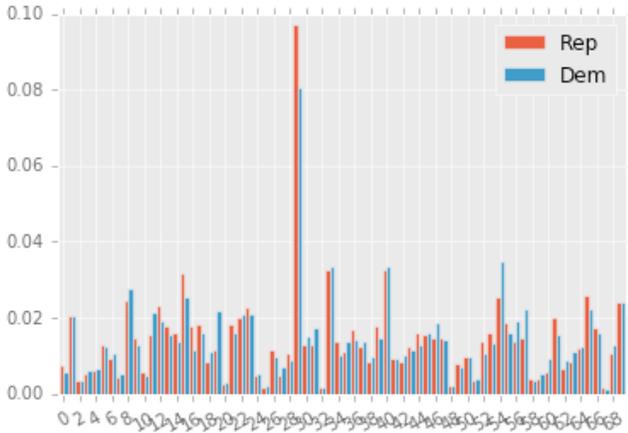

Rep vs. Dem: JS divergence = 0.008

Figure 1. Party Polarization in the U.S. Congress (1994-2016): Rep=Republican Party; Dem=Democratic Party.

The charts in Fig. 2, which corresponds to the Japanese case, compare each of the topic distributions of three different parties (blue bars) with that of the Liberal Democratic Party of Japan (LDP): a long-time ruling, conservative party (red bars). The three other parties are (a) the Democratic Party of Japan (DPJ; center to center-left), (b) the Social Democratic Party (SDP; the Socialist Party before 1996; liberal), and (c) the Japanese Communist Party (JCP; reformist). As is clear from the corresponding values of JS divergence, the pattern of collective attention of any of these parties is considerably different from that of LDP, relative to the difference between Republicans and Democrats in the U.S. Congress. Thus, as far as our measure of polarization is concerned, it can be said that political polarization is markedly smaller in the U.S. than in Japan.

This observation was derived from just a single run of estimation on the underlying topic model for each country in each parameter setting. In order to establish the suggested cross-national difference more rigorously, we iterated the estimation procedure 10 times for each country. The results remain the same. In the case of k=70, for example, the average JS divergence between the Republican and the Democratic parties in the U.S. over 10 runs is 0.00825 with the standard deviation of 0.00071. The corresponding values in the case of Japan are 0.01763±0.00086 (LDP vs. DPJ), 0.02176±0.00144 (LDP vs. SDP), and 0.05125±0.00335 (LDP vs. JCP). Even if one focuses on the relatively small difference between LDP and DPJ, a two-sample Welch's t-test confirms that their JS divergence is still significantly larger than its U.S. counterpart (t-value: 25.197; p-value: 0.000).

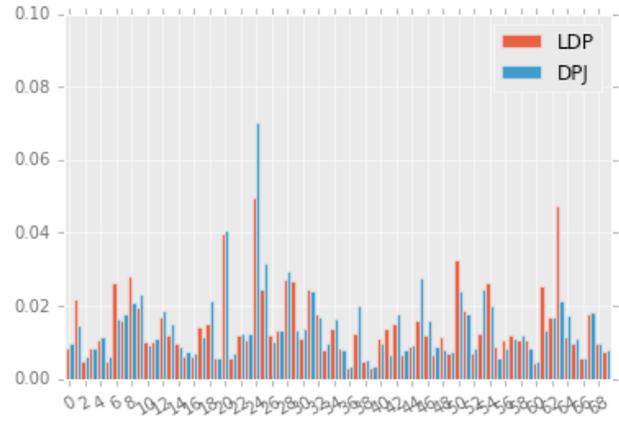

(a) LDP vs. DPJ: JS divergence = 0.018

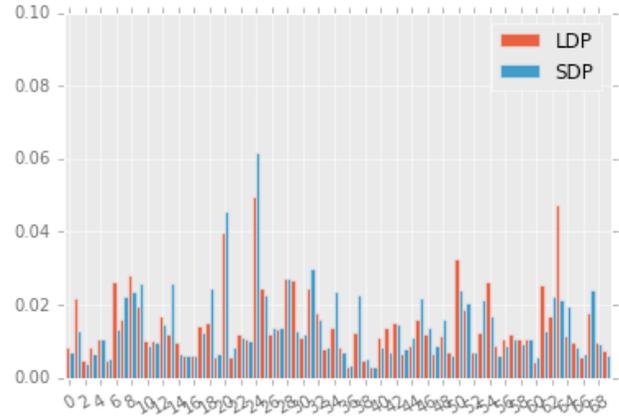

(b) LDP vs. SDP: JS divergence = 0.022

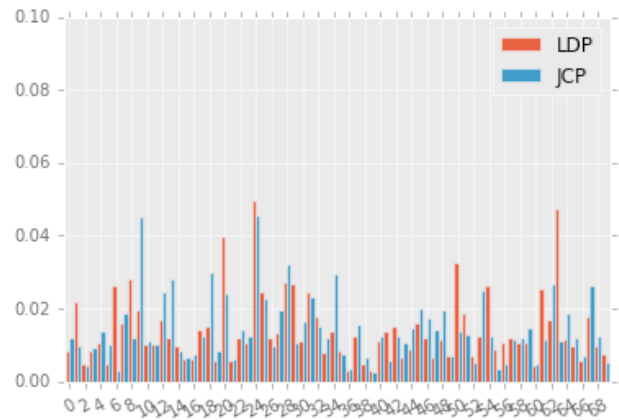

(c) LDP vs. JCP: JS divergence = 0.056

Figure 2. Party Polarization in the Japanese Diet (1994-2016): LDP=Liberal Democratic Party; DPJ=Democratic Party of Japan; SDP=Social Democratic Party; JCP=Japanese Communist Party.

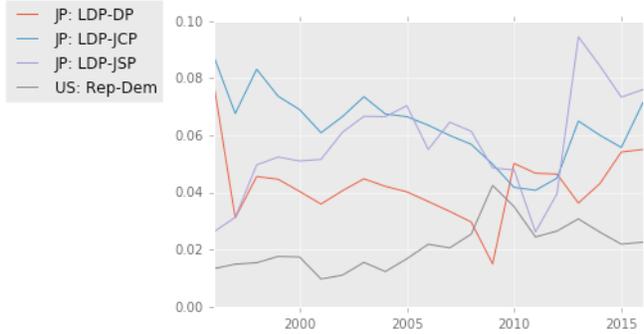

Figure 3. Temporal Dynamics of Political Polization (1994-2016): LDP=Liberal Democratic Party; DPJ=Democratic Party of Japan; SDP=Social Democratic Party; JCP=Japanese Communist Party; Rep=Republican Party; Dem=Democratic Party.

## B. Comparison of Temporal Changes

Fig. 3 plots yearly changes in JS divergences among major political parties of the U.S. and Japan on the same plane. While still retaining the overall cross-national difference in political polarization, the JS divergence of each combination of the parties concerned shows a considerable degree of temporal fluctuation. For example, there is a period of steady rise in inter-party polarization in the U.S. Congress in the latter half of the 2000s. On the other hand, in the Japanese Diet, sharp decline in polarization between LDP and the other parties seems to take place around 2010.

In order to locate possible factors that condition these dynamics, we examined topic-level fluctuations in collective attention of political parties in each country. Fig. 4 illustrates these fluctuations in the U.S. Congress for selected topics that showed distinctive dynamics. Each line plots successive annual inter-party differences in the frequency of reference to the corresponding topic between Republicans and Democrats, that is, for topic $k$, $\Theta_{Dem,Year}(k) - \Theta_{Rep,Year}(k)$. The selected topics are 'health care' (topic 8), 'public spending' (15), 'banks and finance' (35), and 'the wars in Afghanistan and Iraq' (46). These are ideologically divisive issues that dominated the U.S. political scene in the late 2000s for various reasons, including the financial crisis of 2007-2008, the heated controversy surrounding Obamacare (the Affordable Care Act) and the military stalemates in the Middle East. The widening gaps in articulation of these issues between the two parties can be associated with the overall rise in political polarization during the same period.

Figs. 5 and 6 illustrate the corresponding dynamics for two combinations of parties (LDP-DPJ and LDP-JCP) in the case of Japan. The graphs indicate that the Japanese parties had undergone some notable change in their collective attention during the three-year period from 2010 to 2012. This period is exactly one of the unusual political moments when the Democratic Party of Japan (DPJ), rather than the Liberal Democratic Party (LDP), held the government. In Fig. 5, this 'role change' completely reverses the signs of the relative frequencies of reference to some topics such as 'growth strategy' (topic 29) and 'diplomatic negotiation' (61). That is, in this particular period, the DPJ, as the ruling party, was more concerned about these issues than the LDP, whereas otherwise the opposite has always been the case. And during the same period, the LDP, being an opposition, showed a pattern of issue articulation that was relatively similar to those of the other oppositions, including its ideological opposite (the JCP, see Fig.6). The mitigation of political polarization (relative to the LDP) seen in Fig. 3 can be better understood with these structural factors in mind.

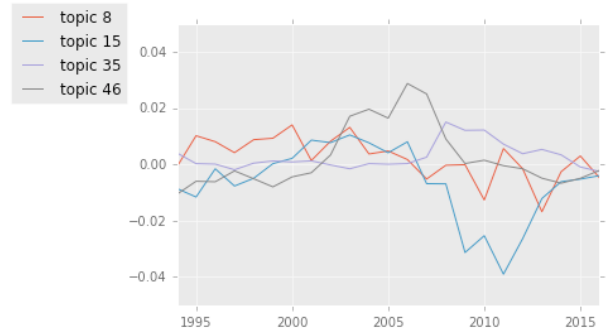

Figure 4. Party Differences in Topic Reference in the U.S. (Dem-Rep): a larger (smaller) value indicates a 'more Democratic (Republican)' topic; topic 8: health care; topic 15: public spending; topic 35: banks and finance; topic 46: wars in Afghanistan and Iraq.

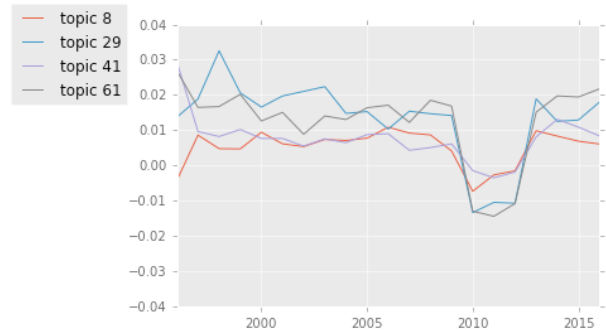

Figure 5. Party Differences in Topic Reference in Japan (LDP-DPJ): a larger (smaller) value indicates a 'more Liberal Democratic (Democratic)' topic; topic 8: foreign policy; topic 29: growth strategy; topic 41: security; topic 61: diplomatic negotiation.

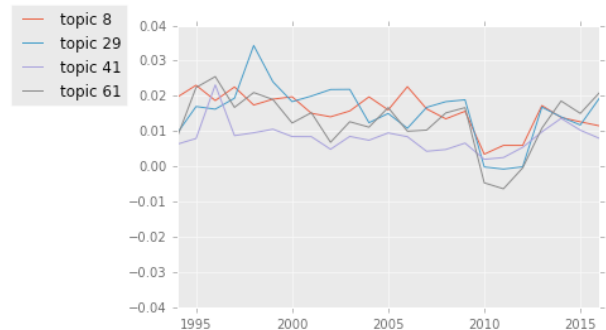

Figure 6. Party Differences in Topic Reference in Japan (LDP-JCP): a larger (smaller) value indicates a 'more Liberal Democratic (Communist)' topic; topic 8: foreign policy; topic 29: growth strategy; topic 41: security; topic 61: diplomatic negotiation.

## IV. Discussion and Conclusions

The comparative analysis performed with the proposed measure of political polarization yielded two findings. First, at least in the past two decades, the Japanese Diet has been more polarized than the U.S. Congress. More specifically, the Japanese political parties have experienced more difficulty in finding 'common grounds' (such as agendas to be set and issues to be discussed) for their public deliberations than their U.S. counterparts. Given the arguments that suggest the 'unusual' nature of the recent political polarization in the U.S. [5], this finding might seem somewhat surprising. However, one should also note a considerable degree of institutional and structural diversity that lies among parliamentary democracies [18,19]. Being a parliamentary cabinet system, the Japanese Diet as a legislature is less autonomous from the executive and more characterized by confrontational deliberations between the governing party and oppositions than the U.S. Congress. The party discipline and restrictions in the Diet are also much stronger and more extensive than those in the Congress, where, without such restrictions, 'cross-voting' across the party boundary has frequently taken place. Finally, the classical argument about the two-party system [20], which suggests the converging tendency of this system to the views of the 'median voter', can be cited here for the relative closeness between Republicans and Democrats in their articulation of political agendas. These factors further illuminate the revealed difference in the degree of political polarization between the two parliamentary democracies.

The same set of factors are also useful for understanding our second finding: the different natures of change in political polarization between the U.S. and Japan. The dramatic reversal in the distribution of attention to several policy topics that accompanied the government change in 2010 is telling evidence for the lack of institutional autonomy of the legislature from the executive in the Japanese context. In contrast, more autonomous, thus more policy-oriented, deliberations in the U.S. Congress are likely to face a severe ideological divide in response to the rise of a prominent policy issue triggered by a specific event (e.g., the global financial crisis of 2007-2008) or a specific act of the executive (e.g., troop deployment to, or withdrawal from, Iraq).

These arguments remain mostly suggestive. More rigorous and more extensive analysis is obviously needed to firmly establish the causal relationships between the revealed patterns of political polarization and the suggested factors. Furthermore, the proposed measure of political polarization should be constantly evaluated for its validity against other comparable measures. As our measure quantifies polarization at a level (i.e., agenda setting and issue articulation) somewhat different from that of most preceding measures (i.e., specific stance or behavior of a lawmaker in given policy issues), the relationships among the measures should also be clarified. Finally, political polarization does not take place on the floor of a legislature alone [4]. It occurs in many different arenas including a society at large (i.e., public opinion). In the future work, we plan to incorporate these broader contexts into our comparative analysis across different countries by employing a wide array of empirical data.

### Acknowledgment

The authors acknowledge valuable financial support from Japan Society for the Promotion of Science (JSPS) (JSPS KAKENHI; Grant Numbers: 16K13347).